# Quantum Mechanical Model for Information Transfer from DNA to Protein

**Ioannis G. Karafyllidis**

*Department of Electrical and Computer Engineering,
Democritus University of Thrace, 67100 Xanthi, Greece
e-mail: ykar@ee.duth.gr*

**Abstract.** A model for the information transfer from DNA to protein using quantum information and computation techniques is presented. DNA is modeled as the sender and proteins are modeled as the receiver of this information. On the DNA side, a 64-dimensional Hilbert space is used to describe the information stored in DNA triplets (codons). A Hamiltonian matrix is constructed for this space, using the 64 possible codons as base states. The eigenvalues of this matrix are not degenerate. The genetic code is degenerate and proteins comprise only 20 different amino acids. Since information is conserved, the information on the protein side is also described by a 64-dimensional Hilbert space, but the eigenvalues of the corresponding Hamiltonian matrix are degenerate. Each amino acid is described by a Hilbert subspace. This change in Hilbert space structure reflects the nature of the processes involved in information transfer from DNA to protein.

# 1. Introduction

Quantum mechanics have been used to explain and describe many phenomena in molecular biology. The importance of the quantum mechanical phenomenon of proton tunneling has been recognized and studied (Lowdin, 1963). Tautomeric mutations are caused by proton tunneling and have been studied extensively (Kryachko, 2002; Douhal et al., 1995). Quantum mechanics have also been used in the study of biomolecule structure and function (Wagner et al., 1998; Home and Chattopadhyaya, 1996; Aspuru-Guzik et al., 2005; Galino et al., 2006; York et al., 1998; Hackermuller et al., 2003). Several attempts have been made to explain the structure of the genetic code using quantum mechanical methods (Eduardo et al., 1993; Balakrishnan, 2002; Forger and Sachse, 2000; Bashford et al., 1998; Frappat et al., 1998). The successful use of quantum mechanics in molecular biology has raised the following question: Can quantum mechanics be used to describe the information storage, processing and transfer by biomolecules? (Davies, 2005; Davies, 2004; Matsuno and Paton, 2000). The work presented in this paper is a contribution towards the construction of a general quantum mechanical model for information processing in biosystems. A model for the information transfer from DNA to protein using quantum information and quantum computation techniques (Nielsen and Chuang, 2000; Karafyllidis, 2003; Karafyllidis, 2004; Karafyllidis, 2005) is developed and presented here.

In living systems information is stored in DNA using a four nucleotide alphabet. This information is copied to mRNA and is transferred to the ribosomes. The tRNA Synthetases recognize the anticodons carried by the tRNA molecules and associate them with the corresponding amino acids. The tRNA molecules reach the ribosomes, where they are associated with the corresponding codon of mRNA and the amino acid they carry binds to the polypeptide chain to form proteins (Wolfe, 1993).

A model for the information transfer from DNA to protein using quantum information and computation techniques is presented in this paper. The model is based on the assumption that genetic information is conserved, and models DNA as the sender and proteins as the receiver of this information. Genetic information for protein synthesis is encoded in nucleotide triplets (codons). Since 64 possible codons exist, a 64-dimensional Hilbert space, for which the codons are the base states, is used to describe the information stored on the sender (DNA) side.

The aim of this work is to device a quantum model for the information transfer from the DNA to protein amino acids, so that the genetic code structure and function can be analyzed using the forthcoming quantum computers. The first and most crucial step in formulating a quantum description a system is to construct its Hamiltonian. Since all codons are different, the corresponding 64X64 Hamiltonian matrix has 64 different (non-degenerate) eigenvalues. The genetic code is degenerate and proteins comprise only 20 different amino acids. Genetic information is conserved and, therefore, the information on the protein side is also described by a 64-dimensional Hilbert space, but the eigenvalues of the corresponding Hamiltonian matrix are degenerate. Each amino acid is described by a Hilbert subspace. The hydrogen bonds of the nucleotides are taken into account in order to raise this degeneracy. The structures of the two Hamiltonian matrices (on sender and on receiver side) are compared and discussed.

In section 2 of this paper an outline of the model and a schematic representation are presented. In section 3 the detailed quantum mechanical model is described, the eigenvalues of the protein Hamiltonian matrix are calculated and the corresponding eigenvectors are determined. The conclusions are presented in section 4.

**2. Model outline**

Genetic information is stored in the DNA molecule using a nucleotide alphabet. There are four elements in this alphabet, the nucleotides, A, C, G and T. The last nucleotide is replaced by U in RNA. DNA stores information for protein synthesis in the form of nucleotide triplets, which are called codons. This information is located in specific parts of the molecule, the genes. According to the central dogma of molecular biology, the information contend of a gene is copied to mRNA molecules, which move and bind to ribosomes. Amino acids and tRNA molecules are assembled together by tRNA Synthetases. The amino acids are carried to the ribosome by the tRNA molecules, which bind to the corresponding codon of mRNA and the amino acid they carry binds to the polypeptide chain to form proteins. Many enzymes and enzymatic reactions are involved in this genetic information transfer.

It is very difficult to describe all the biochemical processes involved in (1) using quantum mechanics and quantum chemistry. A "high-level" description of genetic information processing using quantum information and quantum computation techniques will be developed in this paper. The outline of this description is shown in Figure 1.

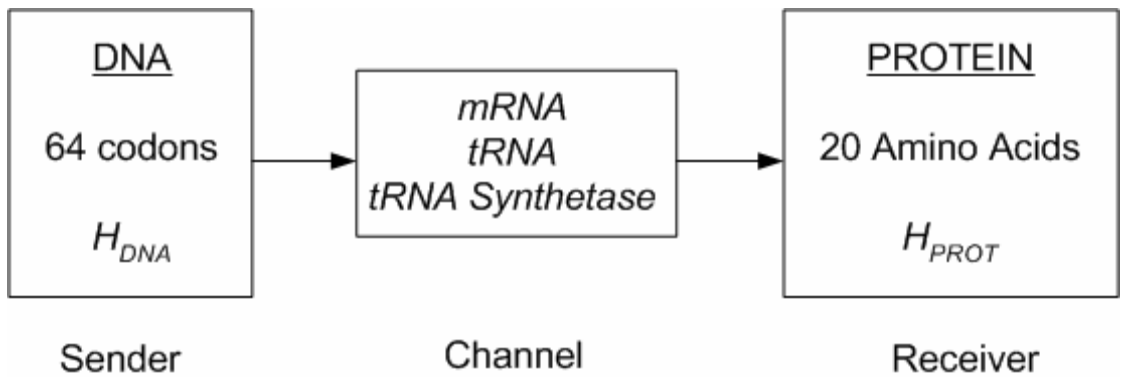

*Figure 1.* Outline of the quantum mechanical model. DNA is the sender of genetic information and protein is the receiver. All the other processes are modeled as the communication channel. $H_{DNA}$ and $H_{PROT}$ are the Hamiltonial matrices on the two sides.

In the quantum mechanical model presented here, DNA is the sender of information, which is stored in the form of codons. The 64 different codons are used to label the 64 base states of the Hilbert space associated with the information stored in DNA. A base state has the form: $|XYZ\rangle$, where X, Y and Z are the nucleotides A, C, G and T. The Hamiltonian matrix for the information contend of DNA, $H_{DNA}$, is:

$$H_{DNA} = \begin{array}{c} \\ |ATG\rangle \\ |TGG\rangle \\ |AAA\rangle \\ |AAG\rangle \\ \vdots \\ |TAA\rangle \\ |TAG\rangle \\ |TGA\rangle \end{array} \begin{array}{c} |ATG\rangle \ |TGG\rangle \ |AAA\rangle \ |AAG\rangle \ \cdots \ |TAA\rangle \ |TAG\rangle \ |TGA\rangle \\ \left[ \begin{matrix} E_{ATG} & 0 & 0 & 0 & \cdots & 0 & 0 & 0 \\ 0 & E_{TGG} & 0 & 0 & \cdots & 0 & 0 & 0 \\ 0 & 0 & E_{AAA} & 0 & \cdots & 0 & 0 & 0 \\ 0 & 0 & 0 & E_{AAG} & \cdots & 0 & 0 & 0 \\ \vdots & \vdots & \vdots & \vdots & \ddots & \vdots & \vdots & \vdots \\ 0 & 0 & 0 & 0 & \cdots & E_{TAA} & 0 & 0 \\ 0 & 0 & 0 & 0 & \cdots & 0 & E_{TAG} & 0 \\ 0 & 0 & 0 & 0 & \cdots & 0 & 0 & E_{TGA} \end{matrix} \right] \end{array} \quad (1)$$

The rows and columns are labeled with the base states. $E_{XYZ}$ is the eigenvalue associated with base state (codon) $|XYZ\rangle$. There are 64 different eigenvalues and since $H_{DNA}$ is diagonal, the base states are also the eigenstates (eigenvectors) of the Hamiltonian.

Information copy from DNA to mRNA, and tRNA-amino acid recognition and matching process, carried out by tRNA Synthetase, are modeled as the communication channel between DNA and protein.

On the protein side, each one of the twenty amino acids is encoded by a set of codons. Two (Met and Trp) are encoded by one codon, nine (Asn, Asp, Cys , Gln, Glu, His, Lys, Phe and Tyr) by two codons, one (Ile) by three, five (Ala, Gly, Pro, Thr and Val) by four and three (Arg, Leu and Ser) by six codons. Three codons (UAA, UAG and UGA) are the stop codons. Taking only this into account the Hamiltonian matrix on this side, $H_{PROT}$, would have only 64 non-zero elements located at its diagonal:

$$H_{PROT} = \begin{array}{c} \\ |AUG\rangle \\ |UGG\rangle \\ |AAA\rangle \\ |AAG\rangle \\ \vdots \\ |UAA\rangle \\ |UAG\rangle \\ |UGA\rangle \end{array} \begin{array}{c} |AUG\rangle \ |UGG\rangle \ |AAA\rangle \ |AAG\rangle \ \cdots \ |UAA\rangle \ |UAG\rangle \ |UGA\rangle \\ \left[ \begin{matrix} E_1 & 0 & 0 & 0 & \cdots & 0 & 0 & 0 \\ 0 & E_2 & 0 & 0 & \cdots & 0 & 0 & 0 \\ 0 & 0 & E_3 & 0 & \cdots & 0 & 0 & 0 \\ 0 & 0 & 0 & E_3 & \cdots & 0 & 0 & 0 \\ \vdots & \vdots & \vdots & \vdots & \ddots & \vdots & \vdots & \vdots \\ 0 & 0 & 0 & 0 & \cdots & E_{S1} & 0 & 0 \\ 0 & 0 & 0 & 0 & \cdots & 0 & E_{S2} & 0 \\ 0 & 0 & 0 & 0 & \cdots & 0 & 0 & E_{S3} \end{matrix} \right] \end{array} \quad (2)$$

The base states are labeled by the codons, but nucleotide T has been replaced by U. These 64 elements are the eigenvalues of the matrix. Two of them, $E_1$ and $E_2$ correspond to the base states $|AUG\rangle$ and $|UGG\rangle$ which encode the amino acids Met and Trp and are not degenerate. $E_3$ corresponds to the base states $|AAA\rangle$ and $|AAG\rangle$ which both encode the amino acid Lys and is two-fold degenerate. Following this line of thought, there are two non-degenerate eigenvalues $E_1$ and $E_2$, corresponding to the codons which are encoding the amino acids Met and Trp, nine two-fold degenerate eigenvalues, $E_3 - E_{11}$ corresponding to the codons encoding the amino acids Asn, Asp, Cys, Gln, Glu, His, Lys, Phe and Tyr, one three-fold degenerate eigenvalue $E_{12}$ corresponding to the codon encoding Ile, five four-fold degenerate eigenvalues $E_{13}$-$E_{17}$ corresponding to the codons encoding the amino acids Ala, Gly, Pro, Thr and Val and three six-fold degenerate eigenvalues $E_{18}$-$E_{20}$ corresponding to the codons encoding the amino acids Arg, Leu and Ser. Eigenvalues $E_{S1}$, $E_{S2}$ and $E_{S3}$ correspond to the three stop codons.

The structure of the Hamiltonian matrices $H_{DNA}$ and $H_{PROT}$ reflects the action of the channel on the information. In the next section a more detailed model for $H_{PROT}$ will be constructed.

## 3. The quantum mechanical model

The Hamiltonian matrix, $H_{PROT}$, which describes the information received on the protein side is in diagonal form and has 64 eigenvalues, from which only five ($E_1$, $E_2$, $E_{S1}$, $E_{S2}$ and $E_{S3}$) are not degenerate. A more detailed description of the information on this side, which is encoded by the amino acids, will be given below.

tRNA molecules comprise an anticodon which is used for their recognition from tRNA Synthetases, which bind the corresponding amino acid to the tRNA molecule. In most of the cases of the anticodon-amino acid recognition and matching process, amino acids are matched with more than one anticodons and hence with more than one codons. The same is true for the anticodon-amino acid assembly and transport process. Only Met and Trp are matched with codons AUG and UGG. During the recognition and matching of a specific anticodon, the recognising and matching system goes to a state that corresponds to this specific anticodon. For example, to recognize the anticodon ACG the system goes to a state we label by $|UGC\rangle$. The same is true for the anticodon-amino acid assembly and transport system. Since a set of anticodons are recognized and matched with the same amino acid, it is reasonable to assume that there are transition probability amplitudes between the corresponding set of states. From now on, transition probability amplitude will be referred to as amplitude. For example, since anticodons UUG and UUA are matched with Asn, it is reasonable to assume that there is an amplitude between the states $|AAC\rangle$ and $|AAU\rangle$. This assumption is incorporated into the Hamiltonian matrix by entering off-diagonal elements. The new form of the Hamiltonian matrix, $H'_{PROT}$, is block-diagonal, with a block corresponding to each amino acid:

$$H'_{PROT} = \begin{bmatrix} E_1 & & & & & & & & \\ & E_2 & & & & & & & \\ & & B_3 & & & & & & \\ & & & \ddots & & & & & \\ & & & & B_{20} & & & & \\ & & & & & E_{S1} & & \\ & & & & & & E_{S2} & \\ & & & & & & & E_{S3} \end{bmatrix} \qquad (3)$$

Blocks $B_3 - B_{20}$ correspond to the 18 amino acids that are encoded by more that one codons. The next step is to determine the off-diagonal elements of $H'_{PROT}$. To do this, we take into account the hydrogen bonds of thee four nucleotides.

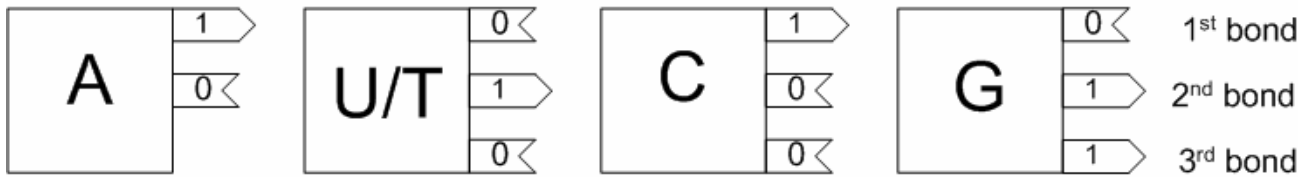

*Figure 2.* Schematic representation of the four nucleotides and their bond sites. Donor sites are marked with "1" and acceptor sites with "0".

Nucleotides C, G and U (T on the DNA side) have three hydrogen bond sites, whereas A has two. Hydrogen bond sites are characterized as "donors" or "acceptors" of hydrogen. Figure 2 shows a schematic representation of the four nucleotides and their bond sites. Donor sites are marked with "1" and acceptor sites with "0". The upper bond is the first bond.

The following three assumptions about the process that will be described are made:

***1. There is no amplitude between any two states corresponding to different amino acids.***

***2. There is an amplitude between two states that correspond to the same amino acid if they differ in one or two continuous nucleotides and have the rest in common.***

For example, for the set of states that correspond to Ser, there is a transition amplitude between $|UCC\rangle$ and $|UCU\rangle$ because they differ only in one nucleotide in the third place. There is also an amplitude between $|UCC\rangle$ and $|AGC\rangle$ because they differ in the two first places, but both have C in the third place. In the set encoding Leu there is no amplitude between $|UUA\rangle$ and $|CUC\rangle$, because they differ in two non-continuous nucleotides (in the first and third place).

***3. Since nucleotide A has only two hydrogen bond sites, we take into account only the first and second bond sites of all nucleotides. If there is an amplitude between two states and the donor/acceptor pattern of the different nucleotides is the same, then the amplitude is ia, where***

***i is the imaginary unit ( $i^2 = -1$ ). If the donors and acceptors are interchanged, then the amplitude is a.***

For example, in the set of states corresponding to Ser, the states $|UCG\rangle$ and $|UCU\rangle$ differ in the third place. The bond pattern of G is (0,1,#) and that of U is (0,1,#), therefore, the amplitude between them is *ia*. In the same set, the states $|UCC\rangle$ and $|AGC\rangle$ differ in the first and second place. The bond pattern of UC is (0,1,#,1,0,#) and that of AG is (1,0,#,0,1,#), therefore, the amplitude between them is *a*. Since *H'$_{PROT}$* must be Hermitian, if the amplitude from $|UCG\rangle$ to $|UCU\rangle$ is *ia*, the amplitude from $|UCU\rangle$ to $|UCG\rangle$ is *–ia*.

Each group of blocks of *H'$_{PROT}$*, will be studied separately.

*3.1 Amino acids encoded by one codon*

Two amino acids are encoded by one codon, namely (Met and Trp), which are encoded by AUG and UGG, respectively. The eigenvalues corresponding to both amino acids are not degenerate neither on DNA or protein side. Although obvious, Table 1 shows the corresponding eigenvalues and eigenvectors on protein side.

***Table 1.*** *Eigenvalues and eigenvectors for amino acids encoded by one codon*

| Amino Acid | Protein side, *H'$_{PROT}$* ||
|---|---|---|
| | Eigenvalue | Eigenvector |
| Met | $E_1$ | $|AUG\rangle$ |
| Trp | $E_2$ | $|UGG\rangle$ |

*3.2 Amino acids encoded by two codons*

There are nine amino acids encoded by two codons, Asn, Asp, Cys , Gln, Glu, His, Lys, Phe and Tyr. For example Asn is encoded by AAC and AAU. Following the three assumptions, described above, the nine blocks in (3) that correspond to these amino acids have the form:

$$B_N = \begin{bmatrix} E_N & a \\ a & E_N \end{bmatrix}, \quad N = 3, 4, 5, \cdots, 11 \tag{4}$$

where N = 3, 4, 5, 6, 7, 8, 9, 10, 11 for Asn, Asp, Cys, Gln, Glu, His, Lys, Phe and Tyr respectively. The eigenvalues and eigenvectors corresponding to blocks $B_3 - B_{11}$, i.e. to these amino acids, on protein side are shown in Table 2.

*Table 2. Eigenvalues and eigenvectors for amino acids encoded by two codons*

| Amino Acid | Protein side, $H'_{PROT}$ | |
|---|---|---|
| | Eigenvalue | Eigenvector |
| Asn | $E_3 - a$ | $(1/\sqrt{2})(|AAC\rangle - |AAU\rangle)$ |
| | $E_3 + a$ | $(1/\sqrt{2})(|AAC\rangle + |AAU\rangle)$ |
| Asp | $E_4 - a$ | $(1/\sqrt{2})(|GAC\rangle - |GAU\rangle)$ |
| | $E_4 + a$ | $(1/\sqrt{2})(|GAC\rangle + |GAU\rangle)$ |
| Cys | $E_5 - a$ | $(1/\sqrt{2})(|UGC\rangle - |UGU\rangle)$ |
| | $E_5 + a$ | $(1/\sqrt{2})(|UGC\rangle + |UGU\rangle)$ |
| Gln | $E_6 - a$ | $(1/\sqrt{2})(|CAA\rangle - |CAG\rangle)$ |
| | $E_6 + a$ | $(1/\sqrt{2})(|CAA\rangle + |CAG\rangle)$ |
| Glu | $E_7 - a$ | $(1/\sqrt{2})(|GAA\rangle - |GAG\rangle)$ |
| | $E_7 + a$ | $(1/\sqrt{2})(|GAA\rangle + |GAG\rangle)$ |
| His | $E_8 - a$ | $(1/\sqrt{2})(|CAC\rangle - |CAU\rangle)$ |
| | $E_8 + a$ | $(1/\sqrt{2})(|CAC\rangle + |CAU\rangle)$ |
| Lys | $E_9 - a$ | $(1/\sqrt{2})(|AAA\rangle - |AAG\rangle)$ |
| | $E_9 + a$ | $(1/\sqrt{2})(|AAA\rangle + |AAG\rangle)$ |
| Phe | $E_{10} - a$ | $(1/\sqrt{2})(|UUC\rangle - |UUU\rangle)$ |
| | $E_{10} + a$ | $(1/\sqrt{2})(|UUC\rangle + |UUU\rangle)$ |
| Tyr | $E_{11} - a$ | $(1/\sqrt{2})(|UAC\rangle - |UAU\rangle)$ |
| | $E_{11} + a$ | $(1/\sqrt{2})(|UAC\rangle + |UAU\rangle)$ |

The eigenvalues associated with each amino acid are not degenerate and the corresponding two eigenvectors are orthogonal and span a 2-dimensional Hilbert subspace. The action of the channel was to associate with each amino acid a 2-dimensional Hilbert subspace.

*3.3 Amino acids encoded by three codons*

There is only one amino acid that is encoded by three codons, Ile. Following the assumptions the block $B_{12}$ is:

$$B_{12} = \begin{array}{c} |AUU\rangle \\ |AUA\rangle \\ |AUC\rangle \end{array} \begin{array}{c} |AUU\rangle \quad |AUA\rangle \quad |AUC\rangle \\ \left[ \begin{array}{ccc} E_{12} & a & a \\ a & E_{12} & ia \\ a & -ia & E_{12} \end{array} \right] \end{array} \quad (5)$$

The rows and columns are labeled with the base states on the DNA side. The eigenvalues and eigenvectors corresponding to this amino acid, on the protein side are shown in Table 3.

*Table 3. Eigenvalues and eigenvectors for the amino acid Ile encoded by three codons*

| Amino Acid | Protein side, $H'_{PROT}$ | |
|---|---|---|
| | Eigenvalue | Eigenvector |
| Ile | $E_{12}$ | $\frac{1}{\sqrt{3}}\left( -i|AUU\rangle - |AUA\rangle + |AUC\rangle \right)$ |
| | $E_{12} - \sqrt{3}\, a$ | $\frac{1}{\sqrt{3}}\left( \frac{1}{2}(i-\sqrt{3})|AUU\rangle - \frac{1}{2}(i\sqrt{3}-1)|AUA\rangle + |AUC\rangle \right)$ |
| | $E_{12} + \sqrt{3}\, a$ | $\frac{1}{\sqrt{3}}\left( \frac{1}{2}(i+\sqrt{3})|AUU\rangle + \frac{1}{2}(i\sqrt{3}+1)|AUA\rangle + |AUC\rangle \right)$ |

The eigenvalues associated with this amino acid are not degenerate and the corresponding three eigenvectors are orthogonal and span a 3-dimensional Hilbert subspace. The action of the channel was to associate this amino acid with a 3-dimensional Hilbert subspace.

*3.4 Amino acids encoded by four codons*

Five amino acids are encoded by four codons, Ala, Gly, Pro, Thr and Val. For example Ala is encoded by GCA, GCC, GCG and GCU.

The five blocks in (3) that correspond to these amino acids have the form:

$$B_M = \begin{bmatrix} E_M & a & ia & a \\ a & E_M & a & ia \\ -ia & a & E_M & a \\ a & -ia & a & E_M \end{bmatrix}, \qquad M = 13, 14, 15, 16, 17 \tag{6}$$

where M = 13, 14, 15, 16, 17 for Ala, Gly, Pro, Thr and Val, respectively. The eigenvalues and eigenvectors corresponding to blocks $B_{13} - B_{17}$, on protein side are shown in Tables 4 - 8.

*Table 4. Eigenvalues and eigenvectors for the amino acid Ala encoded by four codons*

| Amino Acid | Protein side, $H'_{PROT}$ | |
|---|---|---|
| | Eigenvalue | Eigenvector |
| Ala | $E_{13} - (1+\sqrt{2})a$ | $\frac{-1+i}{2\sqrt{2}}|GCA\rangle + \frac{1-i}{2\sqrt{2}}|GCG\rangle - \frac{1}{2}|GCC\rangle + \frac{1}{2}|GCU\rangle$ |
| | $E_{13} + (1-\sqrt{2})a$ | $\frac{-1-i}{2\sqrt{2}}|GCA\rangle + \frac{-1-i}{2\sqrt{2}}|GCG\rangle + \frac{1}{2}|GCC\rangle + \frac{1}{2}|GCU\rangle$ |
| | $E_{13} - (1-\sqrt{2})a$ | $\frac{1-i}{2\sqrt{2}}|GCA\rangle + \frac{-1+i}{2\sqrt{2}}|GCG\rangle - \frac{1}{2}|GCC\rangle + \frac{1}{2}|GCU\rangle$ |
| | $E_{13} + (1+\sqrt{2})a$ | $\frac{1+i}{2\sqrt{2}}|GCA\rangle + \frac{1+i}{2\sqrt{2}}|GCG\rangle + \frac{1}{2}|GCC\rangle + \frac{1}{2}|GCU\rangle$ |

*Table 5. Eigenvalues and eigenvectors for the amino acid Gly encoded by four codons*

| Amino Acid | Protein side, $H'_{PROT}$ | |
|---|---|---|
| | Eigenvalue | Eigenvector |
| Gly | $E_{14} - (1+\sqrt{2})a$ | $\frac{-1+i}{2\sqrt{2}}\|GGA\rangle + \frac{1-i}{2\sqrt{2}}\|GGG\rangle - \frac{1}{2}\|GGC\rangle + \frac{1}{2}\|GGU\rangle$ |
| | $E_{14} + (1-\sqrt{2})a$ | $\frac{-1-i}{2\sqrt{2}}\|GGA\rangle + \frac{-1-i}{2\sqrt{2}}\|GGG\rangle + \frac{1}{2}\|GGC\rangle + \frac{1}{2}\|GGU\rangle$ |
| | $E_{14} - (1-\sqrt{2})a$ | $\frac{1-i}{2\sqrt{2}}\|GGA\rangle + \frac{-1+i}{2\sqrt{2}}\|GGG\rangle - \frac{1}{2}\|GGC\rangle + \frac{1}{2}\|GGU\rangle$ |
| | $E_{14} + (1+\sqrt{2})a$ | $\frac{1+i}{2\sqrt{2}}\|GGA\rangle + \frac{1+i}{2\sqrt{2}}\|GGG\rangle + \frac{1}{2}\|GGC\rangle + \frac{1}{2}\|GGU\rangle$ |

*Table 6. Eigenvalues and eigenvectors for the amino acid Pro encoded by four codons*

| Amino Acid | Protein side, $H'_{PROT}$ | |
|---|---|---|
| | Eigenvalue | Eigenvector |
| Pro | $E_{15} - (1+\sqrt{2})a$ | $\frac{-1+i}{2\sqrt{2}}\|CCA\rangle + \frac{1-i}{2\sqrt{2}}\|CCG\rangle - \frac{1}{2}\|CCC\rangle + \frac{1}{2}\|CCU\rangle$ |
| | $E_{15} + (1-\sqrt{2})a$ | $\frac{-1-i}{2\sqrt{2}}\|CCA\rangle + \frac{-1-i}{2\sqrt{2}}\|CCG\rangle + \frac{1}{2}\|CCC\rangle + \frac{1}{2}\|CCU\rangle$ |
| | $E_{15} - (1-\sqrt{2})a$ | $\frac{1-i}{2\sqrt{2}}\|CCA\rangle + \frac{-1+i}{2\sqrt{2}}\|CCG\rangle - \frac{1}{2}\|CCC\rangle + \frac{1}{2}\|CCU\rangle$ |
| | $E_{15} + (1+\sqrt{2})a$ | $\frac{1+i}{2\sqrt{2}}\|CCA\rangle + \frac{1+i}{2\sqrt{2}}\|CCG\rangle + \frac{1}{2}\|CCC\rangle + \frac{1}{2}\|CCU\rangle$ |

*Table 7. Eigenvalues and eigenvectors for the amino acid Thr encoded by four codons*

| Amino Acid | Protein side, $H'_{PROT}$ | |
|---|---|---|
| | Eigenvalue | Eigenvector |
| Thr | $E_{16} - (1+\sqrt{2})a$ | $\frac{-1+i}{2\sqrt{2}}\|ACA\rangle + \frac{1-i}{2\sqrt{2}}\|ACG\rangle - \frac{1}{2}\|ACC\rangle + \frac{1}{2}\|ACU\rangle$ |
| | $E_{16} + (1-\sqrt{2})a$ | $\frac{-1-i}{2\sqrt{2}}\|ACA\rangle + \frac{-1-i}{2\sqrt{2}}\|ACG\rangle + \frac{1}{2}\|ACC\rangle + \frac{1}{2}\|ACU\rangle$ |
| | $E_{16} - (1-\sqrt{2})a$ | $\frac{1-i}{2\sqrt{2}}\|ACA\rangle + \frac{-1+i}{2\sqrt{2}}\|ACG\rangle - \frac{1}{2}\|ACC\rangle + \frac{1}{2}\|ACU\rangle$ |
| | $E_{16} + (1+\sqrt{2})a$ | $\frac{1+i}{2\sqrt{2}}\|ACA\rangle + \frac{1+i}{2\sqrt{2}}\|ACG\rangle + \frac{1}{2}\|ACC\rangle + \frac{1}{2}\|ACU\rangle$ |

*Table 8. Eigenvalues and eigenvectors for the amino acid Val encoded by four codons*

| Amino Acid | Protein side, $H'_{PROT}$ | |
|---|---|---|
| | Eigenvalue | Eigenvector |
| Val | $E_{17} - (1+\sqrt{2})a$ | $\frac{-1+i}{2\sqrt{2}}\|GUA\rangle + \frac{1-i}{2\sqrt{2}}\|GUG\rangle - \frac{1}{2}\|GUC\rangle + \frac{1}{2}\|GUU\rangle$ |
| | $E_{17} + (1-\sqrt{2})a$ | $\frac{-1-i}{2\sqrt{2}}\|GUA\rangle + \frac{-1-i}{2\sqrt{2}}\|GUG\rangle + \frac{1}{2}\|GUC\rangle + \frac{1}{2}\|GUU\rangle$ |
| | $E_{17} - (1-\sqrt{2})a$ | $\frac{1-i}{2\sqrt{2}}\|GUA\rangle + \frac{-1+i}{2\sqrt{2}}\|GUG\rangle - \frac{1}{2}\|GUC\rangle + \frac{1}{2}\|GUU\rangle$ |
| | $E_{17} + (1+\sqrt{2})a$ | $\frac{1+i}{2\sqrt{2}}\|GUA\rangle + \frac{1+i}{2\sqrt{2}}\|GUG\rangle + \frac{1}{2}\|GUC\rangle + \frac{1}{2}\|GUU\rangle$ |

The eigenvalues associated with each amino acid are not degenerate and the corresponding four eigenvectors are orthogonal and span a 4-dimensional Hilbert subspace. The action of the channel was to associate with each amino acid a 4-dimensional Hilbert subspace.

*3.5 Amino acids encoded by six codons*

Three amino acids are encoded by six codons, Arg, Leu and Ser. Each one will be studied separately.

Arg is encoded by CGA, CGC, CGG, CGU, AGA and AGG. The block $B_{18}$ in (3) that correspond to this amino acid is:

$$B_{18} = \begin{array}{c} \\ |CGA\rangle \\ |CGC\rangle \\ |CGG\rangle \\ |CGU\rangle \\ |AGA\rangle \\ |AGG\rangle \end{array} \begin{array}{c} |CGA\rangle \quad |CGC\rangle \quad |CGG\rangle \quad |CGU\rangle \quad |AGA\rangle \quad |AGG\rangle \\ \begin{bmatrix} E_{18} & ia & a & a & ia & 0 \\ -ia & E_{18} & a & a & 0 & 0 \\ a & a & E_{18} & ia & 0 & ia \\ a & a & -ia & E_{18} & 0 & 0 \\ -ia & 0 & 0 & 0 & E_{18} & a \\ 0 & 0 & -ia & 0 & a & E_{18} \end{bmatrix} \end{array} \quad (7)$$

The rows and columns are labeled with the base states on the DNA side. The eigenvalues and eigenvectors corresponding to this amino acid, on protein side are shown in Table 9.

*Table 9. Eigenvalues and eigenvectors for the amino acid Arg encoded by six codons*

| Amino Acid | Protein side, $H'_{PROT}$ | |
|---|---|---|
| | Eigenvalue | Eigenvector |
| Arg | $E_{18} - a$ | $\frac{1-i}{2\sqrt{3}}|CGC\rangle + \frac{-1+i}{2\sqrt{3}}|CGU\rangle - \frac{1}{\sqrt{3}}|AGA\rangle + \frac{1}{\sqrt{3}}|AGG\rangle$ |
| | $E_{18} + a$ | $-\frac{1+i}{2\sqrt{3}}|CGC\rangle - \frac{1+i}{2\sqrt{3}}|CGU\rangle + \frac{1}{\sqrt{3}}|AGA\rangle + \frac{1}{\sqrt{3}}|AGG\rangle$ |
| | $E_{18} - (1+\sqrt{3})a$ | $\frac{i\sqrt{3}}{\sqrt{12}}|CGA\rangle + \frac{-1+i}{\sqrt{12}}|CGC\rangle + \frac{-i\sqrt{3}}{\sqrt{12}}|CGG\rangle + \frac{1-i}{\sqrt{12}}CGU - \frac{1}{\sqrt{12}}|AGA\rangle + \frac{1}{\sqrt{12}}|AGG\rangle$ |
| | $E_{18} + (1-\sqrt{3})a$ | $\frac{-i\sqrt{3}}{\sqrt{12}}|CGA\rangle + \frac{1+i}{\sqrt{12}}|CGC\rangle + \frac{-i\sqrt{3}}{\sqrt{12}}|CGG\rangle + \frac{1+i}{\sqrt{12}}CGU + \frac{1}{\sqrt{12}}|AGA\rangle + \frac{1}{\sqrt{12}}|AGG\rangle$ |
| | $E_{18} - (1-\sqrt{3})a$ | $\frac{-i\sqrt{3}}{\sqrt{12}}|CGA\rangle + \frac{-1+i}{\sqrt{12}}|CGC\rangle + \frac{i\sqrt{3}}{\sqrt{12}}|CGG\rangle + \frac{1-i}{\sqrt{12}}CGU - \frac{1}{\sqrt{12}}|AGA\rangle + \frac{1}{\sqrt{12}}|AGG\rangle$ |
| | $E_{18} + (1+\sqrt{3})a$ | $\frac{i\sqrt{3}}{\sqrt{12}}|CGA\rangle + \frac{1+i}{\sqrt{12}}|CGC\rangle + \frac{i\sqrt{3}}{\sqrt{12}}|CGG\rangle + \frac{1+i}{\sqrt{12}}CGU + \frac{1}{\sqrt{12}}|AGA\rangle + \frac{1}{\sqrt{12}}|AGG\rangle$ |

Leu is encoded by CUA, CUC, CUG, CUU, UUA, UUG. The block $B_{19}$ in (3) that correspond to this amino acid is:

$$B_{19} = \begin{array}{c} \\ |CUA\rangle \\ |CUC\rangle \\ |CUG\rangle \\ |CUU\rangle \\ |UUA\rangle \\ |UUG\rangle \end{array} \begin{array}{c} |CUA\rangle\ |CUC\rangle\ |CUG\rangle\ |CUU\rangle\ |UUA\rangle\ |UUG\rangle \\ \left[\begin{array}{cccccc} E_{19} & ia & a & a & a & 0 \\ -ia & E_{19} & a & a & 0 & 0 \\ a & a & E_{19} & ia & 0 & a \\ a & a & -ia & E_{19} & 0 & 0 \\ a & 0 & 0 & 0 & E_{19} & a \\ 0 & 0 & a & 0 & a & E_{19} \end{array}\right] \end{array} \qquad (8)$$

The rows and columns are labeled with the base states on the DNA side. The eigenvalues and eigenvectors corresponding to this amino acid, on protein side are shown in Table 10.

*Table 10. Eigenvalues and eigenvectors for the amino acid Leu encoded by six codons*

| Amino Acid | Protein side, $H'_{PROT}$ | |
|---|---|---|
| | Eigenvalue | Eigenvector |
| Leu | $E_{19} - a$ | $\frac{-1-i}{2\sqrt{3}}|CUC\rangle + \frac{1+i}{2\sqrt{3}}|CUU\rangle - \frac{1}{\sqrt{3}}|UUA\rangle + \frac{1}{\sqrt{3}}|UUG\rangle$ |
| | $E_{19} + a$ | $\frac{-1+i}{2\sqrt{3}}|CUC\rangle - \frac{1+i}{2\sqrt{3}}|CUU\rangle + \frac{1}{\sqrt{3}}|UUA\rangle + \frac{1}{\sqrt{3}}|UUG\rangle$ |
| | $E_{19} - (1+\sqrt{3})a$ | $\frac{\sqrt{3}}{\sqrt{12}}|CUA\rangle + \frac{1+i}{\sqrt{12}}|CUC\rangle - \frac{\sqrt{3}}{\sqrt{12}}|CUG\rangle - \frac{1+i}{\sqrt{12}}|CUU\rangle - \frac{1}{\sqrt{12}}|UUA\rangle + \frac{1}{\sqrt{12}}|UUG\rangle$ |
| | $E_{19} + (1-\sqrt{3})a$ | $-\frac{\sqrt{3}}{\sqrt{12}}|CUA\rangle + \frac{1-i}{\sqrt{12}}|CUC\rangle - \frac{\sqrt{3}}{\sqrt{12}}|CUG\rangle + \frac{1-i}{\sqrt{12}}|CUU\rangle + \frac{1}{\sqrt{12}}|UUA\rangle + \frac{1}{\sqrt{12}}|UUG\rangle$ |
| | $E_{19} - (1-\sqrt{3})a$ | $-\frac{\sqrt{3}}{\sqrt{12}}|CUA\rangle + \frac{1+i}{\sqrt{12}}|CUC\rangle + \frac{\sqrt{3}}{\sqrt{12}}|CUG\rangle - \frac{1+i}{\sqrt{12}}|CUU\rangle - \frac{1}{\sqrt{12}}|UUA\rangle + \frac{1}{\sqrt{12}}|UUG\rangle$ |
| | $E_{19} + (1+\sqrt{3})a$ | $\frac{\sqrt{3}}{\sqrt{12}}|CUA\rangle + \frac{1-i}{\sqrt{12}}|CUC\rangle + \frac{\sqrt{3}}{\sqrt{12}}|CUG\rangle + \frac{1-i}{\sqrt{12}}|CUU\rangle + \frac{1}{\sqrt{12}}|UUA\rangle + \frac{1}{\sqrt{12}}|UUG\rangle$ |

Ser is encoded by UCA, UCC, UCG, UCU, AGC and AGU. The block $B_{20}$ in (3) that correspond to this amino acid is:

$$B_{20} = \begin{array}{c} |UCA\rangle \\ |UCC\rangle \\ |UCG\rangle \\ |UCU\rangle \\ |AGC\rangle \\ |AGU\rangle \end{array} \begin{array}{cccccc} |UCA\rangle & |UCC\rangle & |UCG\rangle & |UCU\rangle & |AGC\rangle & |AGU\rangle \end{array} \left[ \begin{array}{cccccc} E_{20} & ia & a & a & 0 & 0 \\ -ia & E_{20} & a & a & a & 0 \\ a & a & E_{20} & ia & 0 & 0 \\ a & a & -ia & E_{20} & 0 & a \\ 0 & a & 0 & 0 & E_{20} & a \\ 0 & 0 & 0 & a & a & E_{20} \end{array} \right] \quad (9)$$

The rows and columns are labeled with the base states on the DNA side. The eigenvalues and eigenvectors corresponding to this amino acid, on protein side are shown in Table 11.

*Table 11. Eigenvalues and eigenvectors for the amino acid Ser encoded by six codons*

| Amino Acid | Protein side, $H'_{PROT}$ | |
|---|---|---|
| | Eigenvalue | Eigenvector |
| Ser | $E_{20} - a$ | $\frac{-1+i}{2\sqrt{3}}\|UCA\rangle + \frac{1-i}{2\sqrt{3}}\|UCG\rangle - \frac{1}{\sqrt{3}}\|AGC\rangle + \frac{1}{\sqrt{3}}\|AGU\rangle$ |
| | $E_{20} + a$ | $-\frac{1+i}{2\sqrt{3}}\|UCA\rangle - \frac{1+i}{2\sqrt{3}}\|UCG\rangle + \frac{1}{\sqrt{3}}\|AGC\rangle + \frac{1}{\sqrt{3}}\|AGU\rangle$ |
| | $E_{20} - (1+\sqrt{3})a$ | $\frac{1-i}{\sqrt{12}}\|UCA\rangle + \frac{\sqrt{3}}{\sqrt{12}}\|UCC\rangle + \frac{-1+i}{\sqrt{12}}\|UCG\rangle - \frac{\sqrt{3}}{\sqrt{12}}\|UCU\rangle - \frac{1}{\sqrt{12}}\|AGC\rangle + \frac{1}{\sqrt{12}}\|AGU\rangle$ |
| | $E_{20} + (1-\sqrt{3})a$ | $\frac{1+i}{\sqrt{12}}\|UCA\rangle - \frac{\sqrt{3}}{\sqrt{12}}\|UCC\rangle + \frac{1+i}{\sqrt{12}}\|UCG\rangle - \frac{\sqrt{3}}{\sqrt{12}}\|UCU\rangle + \frac{1}{\sqrt{12}}\|AGC\rangle + \frac{1}{\sqrt{12}}\|AGU\rangle$ |
| | $E_{20} - (1-\sqrt{3})a$ | $\frac{1-i}{\sqrt{12}}\|UCA\rangle - \frac{\sqrt{3}}{\sqrt{12}}\|UCC\rangle + \frac{-1+i}{\sqrt{12}}\|UCG\rangle + \frac{\sqrt{3}}{\sqrt{12}}\|UCU\rangle - \frac{1}{\sqrt{12}}\|AGC\rangle + \frac{1}{\sqrt{12}}\|AGU\rangle$ |
| | $E_{20} + (1+\sqrt{3})a$ | $\frac{1+i}{\sqrt{12}}\|UCA\rangle + \frac{\sqrt{3}}{\sqrt{12}}\|UCC\rangle + \frac{1+i}{\sqrt{12}}\|UCG\rangle + \frac{\sqrt{3}}{\sqrt{12}}\|UCU\rangle + \frac{1}{\sqrt{12}}\|AGC\rangle + \frac{1}{\sqrt{12}}\|AGU\rangle$ |

The eigenvalues associated with each one of these three amino acids are not degenerate and the corresponding six eigenvectors are orthogonal and span a 6-dimensional Hilbert subspace. The action of the channel was to associate with each amino acid a 6-dimensional Hilbert subspace.

## 4. Discussion

Classical information theory has been extensively used for the analysis and description of biological systems (Yockey, 2005). Furthermore, search and recognition theory has been used to reveal the information processing abilities of biological molecules (Hu and Shklovskii, 2006). It is well known that biological molecules such as DNA and proteins exhibit quantum mechanical properties (Home and Chattopadhyaya, 1996; Hollenberg, 2000). Since these molecules behave quantum-mechanically and since the world is quantum mechanical it is natural to attempt to use quantum information theory to study and describe information transfer between biological molecules.

The aim of this paper is to develop a quantum mechanical model for information transfer from DNA to protein. The steps to develop such a model are to construct the Hamiltonian matrix, calculate is eigenvalues and then calculate its eigenvectors. We followed the Occam's razor principle, i.e. the problem was to find the minimum number of parameters required to construct the Hamiltonian matrix. After several attempts, we finally used only one parameter namely "a", to construct the Hamiltonian of equation (2) from the diagonal (classical) Hamiltonian of equation (1). We also used the donor-acceptor patterns of nucleotide bonds, which have been used for the consideration of the composition of the nucleotide alphabet from the perspective of error-coding theory used in computer science (Mac Dónaill, 2002; Mac Dónaill, 2003).

The main biological consequence of this model is that Hamiltonian matrix has 61 non-degenerate eigenvalues, that correspond to the twenty amino acids, which are encoded by 61 codons. We do not examine the stop codons. A different eigenvalue corresponds to the same amino acid matched and assembled with different anticodons. Therefore, it is possible for the same amino acid to participate in protein synthesis with different molecular energies. Two proteins with exactly the same series of amino acids may have different total energies during synthesis, if the same amino acids are encoded by different codons. If this energy difference is large enough, it may lead to different protein foldings.

## 5. Conclusions

A model for the information transfer from DNA to protein using quantum information and computation techniques was presented in this paper. The model is based on the assumption that genetic information is conserved, and models DNA as the sender and proteins as the receiver of this information. The Hilbert space that describes the genetic information on the sender (DNA) side is spanned by 64 eigenvectors that are labeled with the 64 codons. The biochemical channel, through which the information is transferred, acts on the Hilbert space and separates it into subspaces. The Hilbert space that describes the genetic information on the receiver (protein) side has 18 different subspaces that are associated with the 18 amino acids that are encoded by more than one codon. Each subspace is spanned by the eigenvalues given in Tables 2 – 11. These subspaces are orthogonal to each other. The two amino acids that are encoded by only one codon are associated with the same eigenvalues and eigenvectors in both (DNA and protein) sides. Many questions remain open and the model presented here should be further expanded and elaborated, but this is a first step towards the construction of a complete model for information processing in biosystems, which will be based on quantum information and quantum communication.